# Post-Quantum Discovery as a Governance Capability: Evidence-Based Cryptographic Visibility and Exposure Prioritisation in a Critical Service Provider

Authors: Jelena Zelenovic*[1 2 3], Leila Taghizadeh*[1 2], Edoardo Pena-Gonzalez*[1 2], Jaime Gómez García[2], Bart Preneel*[1]
*Affiliation: COSIC, KU Leuven, Leuven, Belgium

## Abstract

Post-Quantum Cryptography (PQC) readiness is increasingly constrained not by algorithm availability, but by cryptographic visibility, dependency complexity, and fragmented governance. This paper presents an anonymised case study of a large European critical service provider that initiated PQC readiness through a discovery-first strategy, utilizing tool-supported cryptographic inventorying to establish an evidence-based baseline prior to migration planning. The discovery phase revealed systemic challenges, including distributed cryptographic ownership, uneven evidence quality across legacy and modern environments, and high dependency on third-party cryptographic roadmaps. To operationalise these findings, the organisation introduced a structured exposure register that enabled prioritisation based on asset criticality, confidentiality longevity, and migration feasibility. We argue that PQC discovery should be understood as a governance capability that stabilises organisational knowledge and converts cryptographic uncertainty into measurable accountability, supporting risk-based decision-making and ecosystem coordination. The results contribute actionable lessons for institutions pursuing crypto-agility and resilience under post-quantum "harvest-now, decrypt-later" threat models.

## 1 Introduction

Cryptography underpins the security of modern digital infrastructure and is pervasive across communication, business, and information storage systems. A substantial portion of this infrastructure relies on public-key cryptography for key establishment, entity authentication, and digital signatures. However, the anticipated advent of cryptographically relevant quantum computers threatens the security assumptions underlying widely deployed public-key schemes. Consequently, timely migration to Post-Quantum Cryptography is essential to ensure long-term confidentiality, integrity, and trust in digital systems.

The transition to PQC is frequently framed as an algorithm replacement programme: selecting new cryptographic primitives and deploying them across digital infrastructures. However, in large-scale regulated organizations, the main obstacles to PQC readiness are often not by algorithmic selection, but by limited cryptographic visibility, fragmented ownership, and incomplete knowledge as indicated by ENISA, ETSI, and NIST [1- 3], of cryptographic deployment across heterogeneous infrastructure.

In practice, and as discussed by FEDS, NIST, and ETSI [2, 4, 5] organizations struggle with the lack of consolidated knowledge required for planning a PQC migration, including the location of cryptographic applications across infrastructure and applications, end-to-end ownership of cryptographic decisions, identification of assets time-exposed under harvest-now, decrypt-later (HNDL) conditions, assessment of long-lived confidentiality requirements, and visibility into cryptographic dependencies inherited from suppliers and platforms. As a result, migration initiatives are frequently initiated without a baseline of what must be migrated, by whom, or under which operational and dependency constraints.

These knowledge gaps can be summarized as a set of guiding questions that a PQC discovery capability must be able to answer.

- *Where is cryptography used across infrastructure and applications?*
- *Who owns cryptographic decisions end-to-end?*

[1] Contributed equally to this work and is co-first author.
[2] The author writes in a personal and academic capacity. The opinions expressed are solely their own and do not reflect the views or positions of their employer.
[3] Corresponding author: COSIC, KU Leuven, Kasteelpark Arenberg 10 Box 2452, Leuven, B-3001, Belgium. E-mail address: jelena.zelenovic@student.kuleuven.be

- *Which assets are time-exposed under harvest-now, decrypt-later (HNDL) conditions?*
- *Which assets have a long shelf-life and will be relevant to protect at the time that post-quantum cryptography threats enhance?*
- *Which dependencies are internal, and which are inherited from suppliers and platforms?*

**Figure 1.** Guiding questions addressed by PQC discovery as a governance capability.

As per Global Risk Institute and Feder Reserve Board [4, 6], the urgency of addressing this knowledge deficit has increased due to HNDL threats, under which adversaries may capture encrypted traffic now and decrypt it later once cryptographically relevant quantum computers (CRQCs) become available. Under such conditions, and as per Mosca, M. [7] the security of today's encrypted data becomes time-dependent rather than static, and migration feasibility must be evaluated relative to both confidentiality horizons and anticipated threat timelines.

The paper positions PQC discovery as an epistemic precondition for scalable migration [2, 5]. Rather than treating discovery as a preparatory or inventorying task, we argue that discovery functions as an epistemic governance capability: it stabilizes organizational knowledge about cryptographic deployment and dependencies, renders uncertainty explicit, and enables post-quantum exposure to be assessed in a comparable and auditable manner. Without such epistemic stabilization, PQC migration planning risks being speculative, incomplete, or operationally infeasible.

To validate our assertions, we present an anonymized case study of a large European critical service provider that initiated PQC readiness through a discovery-first strategy, employing tool-supported cryptographic inventorying to establish an evidence-based baseline prior to migration planning. The case study examines how discovery outputs were consolidated into a structured exposure register and how this artefact revealed systemic constraints relevant to PQC migration, including ownership fragmentation, uneven evidence quality, and high dependency on third-party cryptographic roadmaps.

This paper investigates whether cryptographic discovery can be operationalized as a governance capability enabling measurable PQC transition readiness in large, regulated organizations.

### 1.1 Contributions

The contributions of this paper are threefold:

1. We propose a discovery-first methodology to establish cryptographic visibility across heterogeneous and distributed technology estates, positioning discovery as a measurable, repeatable, and auditable governance capability rather than a one-time technical inventory exercise.
2. We introduce a structured exposure register model that enables risk-based asset prioritization across multiple dimensions, including systemic criticality, confidentiality longevity, long-term integrity and authentication assurance, and migration feasibility. While confidentiality durability is a primary driver in many post-quantum cryptography (PQC) transition scenarios, we explicitly account for systems characterized by extended operational lifecycles with predominant integrity and authentication requirements. This includes, for example, embedded and field-deployed upgrade or firmware validation mechanisms that may remain operational for 15–20 years or longer, in contexts where post-deployment cryptographic modification is technically constrained or operationally infeasible.
3. We present empirical observations from a large-scale real-world case study, demonstrating that ownership fragmentation, variable evidence confidence levels, and supplier dependency chains represent primary structural barriers to PQC migration and broader crypto-agility adoption.

## 2 PQC Migration and the Need for Crypto-Agility

As discussed across multiple NIST publications [2, 8, 9], PQC readiness is often recognized as a migration challenge rather than a cryptographic operating model. While standardization efforts (e.g., NIST PQC) are progressing, real-world transition depends on organizations' ability to replace cryptographic mechanisms across complex, interdependent systems rather than within isolated components.

In large-scale environments, cryptography is embedded across multiple layers, including application portfolios, middleware and APIs, certificate infrastructures and identity services, legacy platforms with hard-coded cryptographic primitives, externally hosted cloud and SaaS environments, and supplier-managed systems with embedded crypto primitives [2, 3, 5]. PQC transition therefore requires coordinated change across heterogeneous technical and organizational boundaries.

NIST [9] has published initial PQC standards (FIPS 203/204/205) and continues to advance the standardization roadmap, reinforcing that organizations should begin transition planning and implementation preparation. Complementing this effort, the IETF [10] has standardized stateful hash-based signature schemes through the publication of RFCs for XMSS and LMS. However, the availability of standards alone does not guarantee migration feasibility. Organizations frequently lack sufficient visibility and are unable to reliably inventory their cryptographic assets, nor do they possess evidence of sufficient quality to support auditable exposure and risk-based prioritization decisions. As a result, migration planning often begins without a clear understanding of what must be migrated, by whom, and under which technical or operational constraints. Even where organizations have established migration plans, these are in many cases one-off exercises with no embedded strategy for iterative updates or continuous reassessment, which would be required to achieve true crypto-agility as defined by ETSI [3].

These challenges are amplified under HNDL exposure and CRQC horizon planning assumptions that make cryptographic security a function of time. Under such assumptions, time-based exposure models, commonly operationalized using “Mosca inequalities”, are used to identify assets whose confidentiality horizons exceed feasible migration timelines [6, 7]. Applying these models presupposes reliable evidence about cryptographic deployment and dependency structure, which is frequently incomplete or absent in large organizations. Under HNDL threat models, these inequalities provide a practical decision boundary for determining when confidentiality risk becomes structurally unacceptable.

## 3 Discovery Methodology: From Cryptographic Uncertainty to Evidence-Based Exposure

This section describes a discovery methodology that can be used to approximate cryptographic state and migration constraints in a large-scale environment. Within this methodology, an organization may implement a tool-supported discovery pipeline to identify cryptographic assets and dependencies across heterogeneous infrastructure layers.

**Operationalisation of the Quantum Exposure Register (QER)**

To clarify how the Quantum Exposure Register (QER) is constructed and used in practice, this section describes the end-to-end operational workflow linking discovery outputs to governance decision-making.

The QER is not generated as a standalone artefact, but rather emerges from a structured pipeline consisting of four sequential stages:

**1. Data Ingestion from Discovery Pipeline**

Discovery outputs are normalised and ingested into a structured dataset. These inputs originate from:

- Static analysis (e.g., cryptographic libraries, hardcoded keys, configurations)
- Dynamic telemetry (e.g., TLS handshakes, certificates, cipher suites)
- Dependency mapping (e.g., APIs, third-party services, infrastructure layers)

Each detected cryptographic instance is transformed into a candidate **asset record** with the following minimum attributes:

- Asset/service identifier
- Cryptographic mechanism (e.g., RSA-2048, ECC P-256, TLS version)
- System context (application, infrastructure, or vendor dependency)
- Observed usage (e.g., signing, encryption, key exchange)

This step converts raw technical signals into structured, comparable entries.

**2. Enrichment with Governance and Risk Metadata**

Each asset record is then enriched with business and risk-relevant attributes required for prioritisation:

- **Confidentiality horizon (T_shelf):** derived from data classification and retention policies
- **Migration duration (T_migration):** estimated based on system complexity, change windows, and dependencies
- **Ownership (RACI):** mapped to accountable business and technical stakeholders
- **Evidence confidence level:** reflecting quality of discovery data (high/medium/low)
- **Third-party dependency flags:** indicating external constraints

This enrichment step is critical, as it transforms technical inventory into **decision-grade governance data**.

The estimation of temporal variables follows structured governance assumptions. The confidentiality horizon (T_shelf) is derived from data classification policies, regulatory retention requirements, and business-defined data longevity expectations (e.g., financial records, identity data). Migration duration (T_migration) is estimated based on system complexity, dependency chains, change management constraints, and historical transformation benchmarks for comparable systems. The threat horizon (T_threat) is treated as a scenario-based parameter derived from external intelligence sources (e.g., ENISA, GRI, NIST) and is applied consistently across the evaluation cycle to ensure comparability of exposure assessments.
While these variables are subject to uncertainty, their structured estimation enables organizations to convert quantum risk into a comparable and auditable temporal model.

**3. Time-Based Exposure Evaluation**

For each enriched asset, the QER applies the temporal exposure rule defined in Eqn. (1):

An asset is marked as **“time-exposed”** if:

- Its confidentiality lifetime plus migration duration exceeds the assumed threat horizon.

In practice, this is implemented as:

- Automated calculation of exposure status (Yes / Borderline / No)
- Consistent application of a scenario-based **T_threat assumption** across all entries

This step operationalises quantum risk as a **measurable and comparable condition**, rather than an abstract concern.

**4. Prioritisation and Register Population**

The final QER entry is created by combining:

- Exposure status (from Step 3)
- Business criticality (C/I/A classification)
- Evidence confidence

These inputs are then used to:

- Compute a **priority score** (Section 4.2)
- Assign a **migration wave (Wave 1–4)**
- Define **target cryptographic state** (e.g., hybrid, PQC-ready)

- Attach **actionable remediation steps**

The result is a structured register entry that links:

**technical cryptographic reality → time-based risk → accountable action**

**5. Continuous Update and Governance Integration**

The QER is not static; it is maintained as a **living governance instrument**:

- Updated periodically as discovery coverage improves
- Refreshed when threat horizon assumptions change
- Integrated into:
  - Risk committees
  - Architecture governance
  - Third-party risk processes

This ensures that PQC readiness evolves from a one-off exercise into a continuous control process.

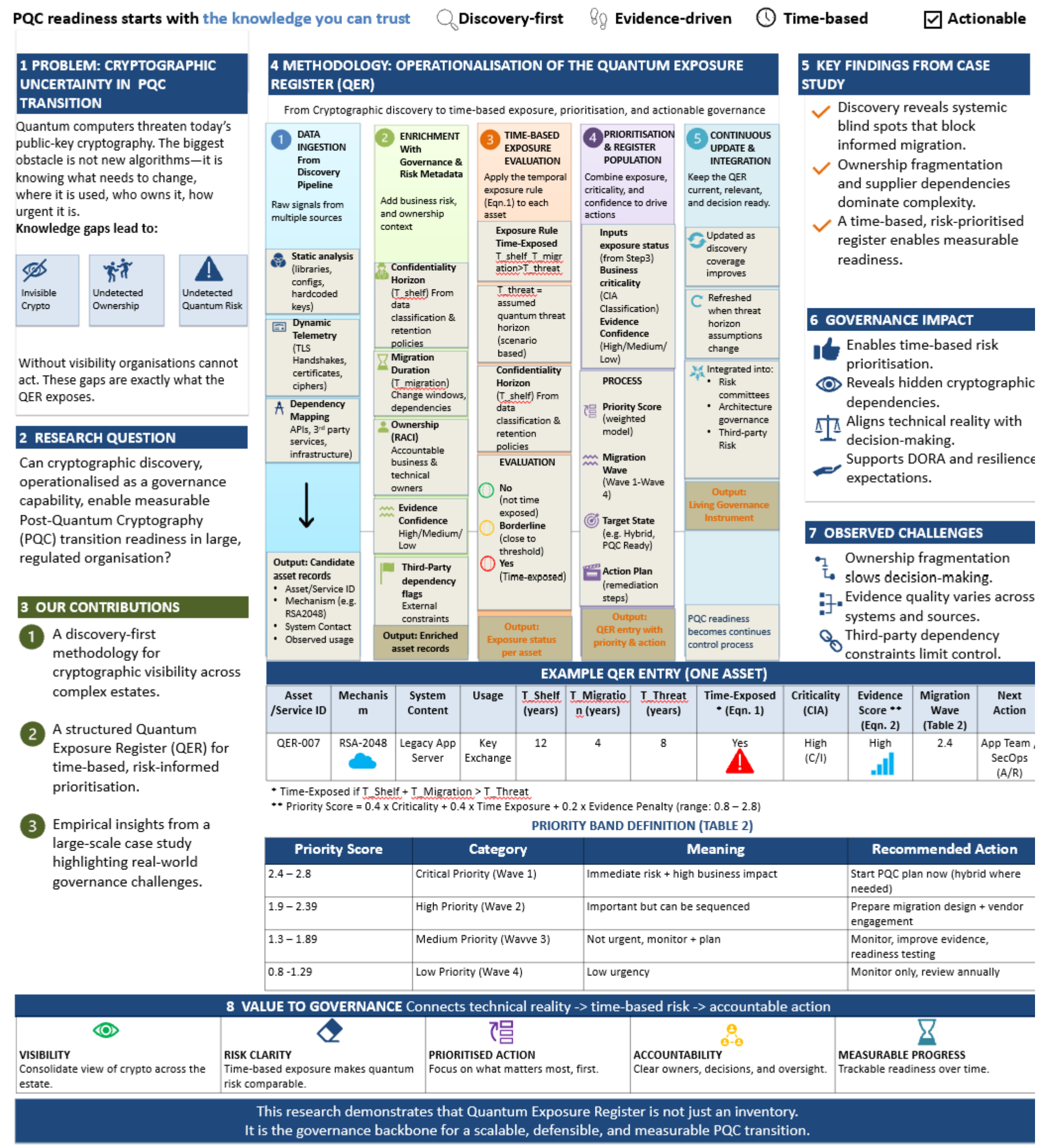

| Asset /Service ID | Mechanism | System Content | Usage | T_Shelf (years) | T_Migration (years) | T_Threat (years) | Time-Exposed * (Eqn. 1) | Criticality (CIA) | Evidence Score ** (Eqn. 2) | Migration Wave (Table 2) | Next Action |
|---|---|---|---|---|---|---|---|---|---|---|---|
| QER-007 | RSA-2048 | Legacy App Server | Key Exchange | 12 | 4 | 8 | Yes | High (C/I) | High | 2.4 | App Team , SecOps (A/R) |

| Priority Score | Category | Meaning | Recommended Action |
|---|---|---|---|
| 2.4 – 2.8 | Critical Priority (Wave 1) | Immediate risk + high business impact | Start PQC plan now (hybrid where needed) |
| 1.9 – 2.39 | High Priority (Wave 2) | Important but can be sequenced | Prepare migration design + vendor engagement |
| 1.3 – 1.89 | Medium Priority (Wavve 3) | Not urgent, monitor + plan | Monitor, improve evidence, readiness testing |
| 0.8 -1.29 | Low Priority (Wave 4) | Low urgency | Monitor only, review annually |

The approach combines automated tools for scanning and inventorying cryptographic protocols, certificates, key sizes, and algorithm usage; dependency mapping across applications, APIs, networks, cloud services, and third-party integrations; and confidentiality horizon classification linking encrypted assets to data longevity requirements and regulatory retention obligations [2, 5, 11].

Furthermore, assets are flagged as quantum-exposed when confidentiality requirements outlasted feasible migration timelines. In this formulation, discovery is not merely mapping; it becomes temporal risk diagnosis that converts uncertainty into a governance prioritisation list [4-7].

Formally, as described in Eqn. (1) below, an asset $A_i$ is treated as quantum-exposed if [6,7]:

$$T_{shelf}(A_i) + T_{migration}(A_i) > T_{threat} \quad , \qquad (1)$$

where:

- $T_{shelf}(A_i)$= required confidentiality lifespan ("confidentiality horizon")
- $T_{migration}(A_i)$= migration time required (including dependencies and change windows)
- $T_{threat}$= threat horizon (time to CRQC relevance for targeted cryptosystems)

The exposure rule applied in Eqn. (1) is conceptually aligned with Mosca's inequality [7], which defines quantum risk as a function of the relationship between data lifetime, migration time, and adversarial capability timelines. While Eqn. (1) is presented in simplified form for operational use, it represents a direct instantiation of this inequality within an enterprise governance context.
The automated discovery process is implemented through a combination of static analysis, dynamic telemetry inspection, and dependency mapping. Static discovery leverages source code analysis tools (e.g., SAST frameworks and cryptographic scanners) to identify cryptographic library usage, hardcoded keys, and algorithm references across application repositories. Dynamic discovery complements this by capturing runtime cryptographic behaviour through TLS handshake inspection, certificate parsing, and network telemetry analysis in production environments. Dependency mapping further extends discovery to APIs, third-party services, and infrastructure layers, enabling identification of externally inherited cryptographic mechanisms.
This multi-layered approach ensures that discovery outputs reflect both intended cryptographic configurations and actual operational behaviour, thereby increasing evidence completeness and reducing false negatives.

## 4 Discovery as the Enabler of Crypto-Agility

This case study suggests that discovery-first approaches provide a practical foundation for crypto-agility by reducing uncertainty about cryptographic deployment, ownership, and dependency structure. Rather than accelerating migration directly, discovery enables organizations to reason about migration feasibility in a repeatable and evidence-based manner.

### 4.1 Why Discovery is a Governance Capability

Discovery functions as a governance capability by producing a measurable and auditable baseline of cryptographic state. Ownership mapping supports accountability, evidence confidence scoring makes uncertainty explicit, and temporal exposure logic enables prioritization based on confidentiality longevity and migration feasibility. Dependency visibility further allows supplier constraints to be incorporated into planning assumptions. In this sense, discovery does not constitute a migration plan, but establishes the preconditions required for migration planning to be defensible and repeatable.
In this sense, discovery produces governance artefacts analogous to other resilience diagnostics: not a final migration plan, but the precondition for one.

### 4.2 Implications for Critical Service Providers and Financial Institutions

The approach is particularly relevant for organizations with long-lived confidentiality requirements, deep legacy dependencies, and ecosystem-driven cryptographic inheritance. For regulated critical service providers and financial institutions, discovery-first approaches support crypto-agility by stabilizing cryptographic knowledge and enabling exposure to be assessed and communicated under resilience and control assurance expectations.

### 4.3 Operational Methodologies for PQC Transition

**Dual-Layer Cryptographic Discovery Protocol (Static and Dynamic Evidence Acquisition)**

A primary challenge in enterprise-scale cryptographic transition programmes is the presence of "hidden cryptography" and heterogeneous evidence quality across legacy and modern environments. To address these limitations, organizations should adopt a dual-layer discovery protocol combining static analysis techniques with runtime telemetry validation.

Static discovery focuses on identifying cryptographic dependencies embedded within application codebases, configuration artefacts, and infrastructure templates [11]. In practice, this includes static application security testing (SAST) tuned specifically for cryptographic function calls (e.g., OpenSSL, Bouncy Castle, platform-native cryptographic libraries), as well as detection of hard-coded keys, deprecated algorithms, and insecure key management patterns across source code and configuration repositories. Static analysis provides architectural visibility and supports early identification of migration complexity drivers.

Dynamic discovery complements static techniques by validating cryptographic behaviour under real operational conditions. Runtime telemetry and network-level inspection enable capture of Transport Layer Security (TLS) handshakes, negotiated cipher suites, and certificate exchange behaviour in production environments [10]. This ensures that cryptographic discovery reflects actual system behaviour rather than intended configuration states. The combination of static and dynamic discovery substantially improves evidence confidence and reduces both false-positive and false-negative exposure classification [10, 11].

**Temporal Risk Prioritization and Time-Bounded Exposure Modelling**

Traditional asset prioritization models based solely on business criticality are insufficient for post-quantum transition planning. Cryptographic exposure must instead be evaluated as a time-dependent risk variable, incorporating both confidentiality longevity requirements and migration feasibility constraints.

Temporal prioritization can be operationalized using Mosca-type exposure reasoning, whereby an asset is considered quantum-exposed if the sum of its required confidentiality lifetime and estimated migration duration exceeds the projected threat horizon for a cryptographically relevant quantum capability [4, 6, 7]. Recent quantum-safe financial risk modelling research further reinforces the need to integrate time-based cryptographic exposure into enterprise risk decision frameworks [4].

In practice, organisations should implement:

- Confidentiality longevity classification, assigning each cryptographic asset a required trust lifetime.
- Migration feasibility scoring, incorporating technical complexity, third-party dependency constraints, operational change windows, and resource availability.
- Priority escalation for assets combining long confidentiality horizons with high migration complexity, which represent structurally constrained transition risks.

**Dynamic Quantum Exposure Register (QER) as a Governance Control Instrument**

The Quantum Exposure Register (QER) should be implemented as a continuously updated governance artefact integrated into enterprise risk management (ERM) processes [2, 3, 5]. Rather than operating as a static inventory, the register functions as a decision-support system linking technical discovery outputs to governance accountability and remediation planning [2, 12]. By embedding Eqn. (1) into exposure registers and reporting workflows, cryptographic risk becomes measurable and auditable at enterprise level.

Each register entry should, at minimum, document the currently deployed cryptographic algorithm or primitive; the designated owner according to the RACI accountability model; a quantified crypto-agility score reflecting the relative ease and operational feasibility of algorithm replacement; and a defined remediation deadline derived from a combined assessment of exposure risk and migration feasibility constraints.

From an operational governance perspective, the QER becomes the authoritative source for executive-level reporting. It enables consistent exposure prioritization, audit traceability, and cross-functional coordination across security engineering, resilience governance, procurement, and third-party risk management functions.

**Supply Chain Cryptographic Governance Through Crypto-SBOM Integration**

Given the increasing externalization of cryptographic functionality through SaaS platforms, managed services, and embedded vendor components, PQC readiness must be extended beyond internal infrastructure boundaries into supplier ecosystems.

To reduce third-party cryptographic opacity, organizations should extend software supply chain transparency principles through Crypto-Software Bills of Materials (Crypto-SBOMs) [14, 15, 16]. These artefacts should document cryptographic primitives, key management dependencies, certificate lifecycle mechanisms, and algorithm implementation locations within vendor-delivered systems.

Procurement and supplier governance functions should incorporate [2, 14]:

- Mandatory Crypto-SBOM disclosure requirements for new procurements and major upgrades.
- Contractual clauses requiring vendor PQC transition roadmaps and disclosure of cryptographic dependency changes.
- Time-bounded cryptographic readiness assurance checkpoints (e.g., annual or 12-month review cycles).

This approach shifts PQC readiness from a purely internal engineering programme to a coordinated ecosystem-level transition strategy.

**Cryptographic Abstraction and Hybrid Transition Architectures**

To minimize repeated migration cost and reduce transition risk under evolving PQC standards, organizations should design cryptographic agility directly into system architectures. This is achieved through abstraction-layer design patterns that decouple application logic from specific cryptographic implementations [2, 3].

Abstraction mechanisms may include cryptographic service wrappers, interface-driven crypto libraries, and centralized cryptographic service platforms. These design patterns enable algorithm substitution without requiring full application redesign, significantly reducing transition friction during PQC standard maturation.

During transition periods, hybrid cryptographic deployment strategies should be considered, combining classical cryptography with PQC primitives to maintain backward compatibility while introducing forward-resistant protections [19, 20]. Hybrid implementations reduce transition risk during standardization uncertainty phases and support progressive migration strategies aligned with supplier readiness and interoperability constraints.

## 5 Exposure Register as a Governance Instrument

Discovery outputs were consolidated into an internal structured register comparable to a Quantum Exposure Register (QER). The register enables comparison and prioritization of assets based on criticality, confidentiality longevity, technical migration feasibility, and supplier dependency constraints. The QER is a governance artefact that translates cryptographic discovery outputs into an actionable, auditable, time-based prioritization list, enabling the organization to manage HNDL exposure and plan PQC migration sequencing [2, 3]. The Quantum Exposure Register operationalizes Eqn. (1) by transforming temporal exposure into an auditable asset-level governance register.

### 5.1 Case Study: Quantum Exposure Register (QER) in Organisation A (Anonymised)

**Organization Profile:** Organization A is a large European critical service provider operating at the intersection of postal logistics, digital identity, financial services, and payments. The organization runs a high-volume transactional environment combining legacy infrastructure, modern cloud-enabled services, customer-facing digital channels, and extensive third-party service interconnections. The institution initiated PQC readiness with the objective of reducing exposure to long-lived confidentiality domains and addressing HNDL threats.

**Implementation Rationale:** Rather than starting directly with algorithm substitution, Organization A adopted a discovery-first approach – utilizing the methodology described in Sect. 3 – and specifying the fields in the QER (which we refer as a recommended minimum set). Following discovery, the organization consolidated outputs into the QER. Each entry connected cryptographic assets to confidentiality horizons and migration feasibility. The QER enabled time-based prioritization using the inequality (1), while preserving auditability through evidence confidence scoring and explicit RACI (Responsible, Accountable, Consulted, and Informed) ownership. This structure allowed the organization to isolate a small subset of highly time-exposed critical services (Wave 1) and to differentiate internal migration actions from supplier-dependent transition constraints.

As shown in Table 1, Organization A established a minimum baseline schema for the Quantum Exposure Register (QER), under which each register entry captures the following attributes: service or asset identifier; associated business function or domain; criticality classification across confidentiality, integrity, and availability (C/I/A);

required data confidentiality horizon (T_shelf); estimated migration duration (T_migration); scenario-based threat horizon assumption (T_threat), maintained consistently across the register evaluation cycle; time-exposure status derived from the applicable temporal risk inequality (1); currently deployed cryptographic mechanisms or protocols; evidence confidence level (e.g., high, medium, low); ownership allocation based on RACI principles, covering both business and technical accountability; third-party dependency constraints; a quantified crypto-agility score (1–5 scale); defined target cryptographic end state (e.g., hybrid[4], post-quantum, or supplier-led transition); and the corresponding remediation action plan, including defined deadlines or migration wave allocation.

**Organisation A utilised Time-exposed rule parameters, as per Eqn. (1):** Assumed threat horizon for this cycle: $T_{threat} = 8$ years (example planning assumption used consistently across the register cycle).

**Table 1.** Case Study (Organisation A) QER implementation's extract (sample entries, anonymised) [4, 5, 12, 13].

[4] In this context, "hybrid" denotes deployment states in which classical and post-quantum cryptographic mechanisms are operated concurrently to support phased migration, interoperability requirements, and risk-managed transition toward a PQC-native end state.

| QER ID | Asset / Service (anonymised) | Domain | Criticality (C/I/A) | T_shelf (yrs) | T_migration (yrs) | T_threat (yrs) | Time-Exposed? | Current Crypto | Evidence Confidence | Owner (Biz / Tech) | Third-Party Dependency | Crypto-Agility (1–5) | Target State | Next Action | Wave |
|---|---|---|---|---|---|---|---|---|---|---|---|---|---|---|---|
| QER-001 | National Digital Identity Signing Service | Identity / Trust | High / High / Med | 15 | 3 | 8 | Yes | RSA-2048 signatures, TLS | High | Identity Lead / PKI Lead | HSM vendor, CA chain | 2 | Hybrid → PQC signatures | Crypto-SBOM [14, 15, 16] + signing algorithm upgrade plan | Wave 1 |
| QER-002 | Payment Transaction Gateway | Payments | High / High / High | 10 | 4 | 8 | Yes | ECC + RSA mix, mutual TLS | High | Payments Exec / AppSec Lead | External payment rails | 2 | Hybrid TLS + PQC readiness | Vendor PQC roadmap + test environment | Wave 1 |
| QER-004 | Interbank Messaging Interface | Banking Connectivity | High / High / High | 12 | 5 | 8 | Yes | RSA TLS + legacy cipher suites | Med | Banking Ops / Network Sec | SWIFT-like dependency | 1 | Hybrid + segmentation | Immediate cipher suite hardening + PQC transition assessment | Wave 1 |
| QER-005 | Public Key Infrastructure (PKI) Core | Foundational | High / High / High | 15 | 6 | 8 | Yes | RSA root/intermediates | High | CIO Sponsor / PKI Lead | CA vendor, tooling | 1 | PQC-enabled PKI design | Design future PKI hierarchy + pilot PQC cert profiles | Wave 1 |
| QER-009 | Customer KYC Vault | Customer Data | High / High / Med | 15 | 4 | 8 | Yes | RSA key wrapping + TLS | High | Chief Data Owner / Sec Eng | DLP provider | 2 | PQC migration target | Prioritise vault + re-encryption feasibility study | Wave 1 |
| QER-011 | Legacy Mainframe Batch Settlement | Legacy Core | High / High / High | 10 | 7 | 8 | Yes | RSA embedded + custom crypto | Low | Core Banking / Legacy Team | Vendor unclear | 1 | Compensating controls + plan | Immediate owner assignment + discovery deep dive | Wave 1 |
| QER-007 | Logistics IoT Device Fleet (sorting centers) | Operations | High / Med / High | 10 | 6 | 8 | Yes | Embedded RSA + VPN tunnels | Low | Ops Owner / OT Security | Device manufacturer | 1 | Replace/refresh + PQC capable | Create replacement schedule + procurement crypto requirements | Wave 2 |

| QER-008 | Regulatory Reporting Archive Repository | Compliance | High / High / Low | 12 | 2 | 8 | Yes | Encrypted storage + RSA keys | Med | Compliance Head / Data Security | Storage vendor | 3 | Hybrid encryption model | Identify long-life data + key rotation plan + PQC readiness | Wave 2 |
|---|---|---|---|---|---|---|---|---|---|---|---|---|---|---|---|
| QER-010 | API Integration Layer (partner ecosystem) | External Integration | High / High / Med | 8 | 3 | 8 | Borderline | ECC TLS + partner certs | Med | Partnerships / API Lead | External partners | 3 | Hybrid TLS | Partner readiness survey + staged rollout plan | Wave 2 |
| QER-012 | SOC / SIEM Log Integrity & Signing | Security Ops | Med / High / High | 10 | 2 | 8 | Yes | RSA signing / hashing | High | CISO / SOC Lead | SIEM vendor | 4 | Hybrid signing | Introduce PQC-ready signing option for forensic integrity | Wave 2 |
| QER-003 | Citizen Portal (Customer-facing TLS) | Digital Channels | Med / High / Med | 5 | 2 | 8 | No | ECC TLS (P-256) | High | Digital Owner / Cloud Sec | CDN provider | 4 | PQC-hybrid TLS when supported | Monitor supplier support + client impact analysis | Wave 3 |
| QER-006 | Internal HR & Payroll SaaS | Admin SaaS | Med / Med / Med | 7 | 1 | 8 | No | SaaS-managed TLS | Low | HR Owner / Vendor Mgmt | Fully SaaS-managed | 3 | Supplier-led | Obtain assurance statement + contract clause update | Wave 4 |

### 5.2 QER Scoring Model Definition

To support consistent prioritisation, a quantitative scoring model for entries in the Quantum Exposure Register can be devised [13]. The proposed model computes a composite priority score as the weighted sum of three independent components: asset criticality, time-based exposure status, and evidence quality. Each component contributes separately to the overall score and is assigned a fixed weight (0.4, 0.4, and 0.2, respectively), reflecting equal emphasis on business impact and exposure window, with evidence quality acting as a secondary modifier rather than a primary driver, consistent with impact-likelihood reasoning adapted for PQC transition risk.

- The three components are defined as follows:
- **Criticality score:** this score refers to business impact, where High=3, Med=2, Low=1
- **Time Exposure Score:** this score refers to the urgency and the exploit window related to likelihood and attack timing, where Yes=3, Borderline=2, No=1
- **Evidence Confidence Penalty:** this score reflects the data quality and uncertainty with High=0, Med=1, Low=2 *(higher penalty if evidence weak)*

The overall formula is then considered as shown below in Eqn. (2):

$$Priority = (CriticalityScore \times 0.4) + (TimeExposureScore \times 0.4) + (EvidenceConfidencePenalty \times 0.2) \quad (2)$$

This scoring approach is conceptually aligned with established risk modelling and operational resilience frameworks, including:

- **FAIR-style quantitative risk thinking**, where risk is derived from loss magnitude and loss event frequency, and where uncertainty and data quality materially influence risk estimation confidence [17]
- **NIST risk modelling approaches**, particularly those emphasising likelihood, impact, and confidence in available evidence when supporting risk-informed decision making [18].
- **DORA-aligned operational risk logic**, which implicitly combines impact severity, threat/exposure conditions, and uncertainty when assessing ICT-related risk scenarios and resilience posture [12].

The weighting scheme (0.4, 0.4, 0.2) reflects an intentional prioritisation logic grounded in operational risk modelling principles. Criticality and time exposure are assigned equal weight as primary drivers of impact and urgency, respectively, consistent with impact-likelihood paradigms in frameworks such as FAIR. Evidence confidence is treated as a secondary modifier to avoid over-penalising assets where uncertainty exists, while still capturing the governance risk associated with incomplete visibility.

Furthermore, as shown in Table 2, the calculated priority scores are mapped to discrete priority bands that translate numerical results into operational PQC migration waves. These bands represent organization-specific thresholds used to sequence migration activities, rather than absolute measures of cryptographic risk. Table 2 therefore defines an interpretation model that organizations can adapt to align scoring outputs with concrete migration actions.

**Table 2.** Priority Band Definition and Interpretation.

| Priority Score | Category | Meaning | Recommended Action |
|---|---|---|---|
| 2.4 – 2.8 | Critical Priority (Wave 1) | Immediate risk + high business impact | Start PQC plan now (hybrid where needed) |
| 1.9 – 2.39 | High Priority (Wave 2) | Important but can be sequenced | Prepare migration design + vendor engagement |
| 1.3 – 1.89 | Medium Priority (Wave 3) | Not urgent, monitor + plan | Monitor, improve evidence, readiness testing |

| 0.8 – 1.29 | Low Priority (Wave 4) | Low urgency | Monitor only, review annually |
|---|---|---|---|

The priority band thresholds defined in Table 2 are organisation-specific calibration parameters derived from internal risk tolerance and operational capacity constraints. Rather than representing absolute measures of cryptographic risk, these bands function as decision thresholds that translate continuous priority scores into actionable migration waves. In the case study organisation, thresholds were calibrated through iterative validation workshops involving security, risk, and architecture stakeholders, ensuring alignment between model outputs and practical execution capacity.

Assignment of assets to migration waves is not purely automated but follows a hybrid decision model. Initial wave allocation is generated algorithmically based on the computed priority score and associated threshold bands, after which allocations are reviewed through governance processes (e.g., risk committees, architecture boards) to account for operational constraints, interdependencies, and supplier readiness. This hybrid approach ensures both consistency and practical feasibility in migration sequencing.

### 4.3 QER Priority Score Implementation Example (1 anonymised asset)

The following example demonstrates application of the scoring approach defined in Section 4.2.

**Example Asset:** QER-009 — Customer KYC Vault

**1) Step: Assign scores**

**Criticality (C/I/A):** High / High / Medium
**Criticality Score** = 3

**Time-Exposed?** Yes
**Time Exposure Score** = 3

**Evidence Confidence:** High
**Evidence Confidence Penalty** = 0 *(because High evidence = no penalty)*

**2) Step: Apply the formula in Eqn. (2)**

**Substitute values:**

$$Priority = (3 \times 0.4) + (3 \times 0.4) + (0 \times 0.2)$$
$$Priority = 1.2 + 1.2 + 0$$

**Final Priority Score**

$$Priority = 2.4$$

**Interpretation:**

Given the defined priority bands in Table 2, a Priority Score = 2.4 is very high, meaning this service should be treated as Wave 1 (urgent) because:

- it is highly critical to the organisation
  it is time-exposed under HNDL
  → the evidence is strong enough to act immediately

## 6 Case Study Findings: Challenges Observed During PQC Discovery at Organisation A

In the early stages of discovery phase Organization A was able to identify the following empirical observations that underlined persistent governance and operational challenges.

### 6.1 Fragmented Cryptographic Ownership (“Hidden Crypto”)

A dominant finding was that encryption and key management responsibilities were distributed across multiple teams and platforms (application engineering, infrastructure operations, cloud services, and vendor-managed services). This fragmentation created "hidden crypto" zones, where cryptographic decisions were embedded in libraries, vendor components, and legacy middleware, reducing traceability and slowing readiness progression.

### 6.2 Uneven Evidence Quality Across Environments

Discovery evidence quality varied by system maturity. Modern workloads provided richer telemetry and certificate visibility, while legacy systems produced partial or indirect evidence. To preserve auditability, the organization introduced an evidence confidence scoring approach (e.g., High/Medium/Low) to distinguish reliable findings from inferred findings.

### 6.3 Third-Party Dependencies and Cryptographic Inheritance Risk

Organization A reported high dependency on supplier roadmaps, especially where SaaS providers or payment platforms encapsulated cryptography behind contractual interfaces. Discovery thus became inseparable from third-party assurance, requiring the organization to treat PQC readiness as an ecosystem dependency rather than an isolated internal programme.

### 6.4 Completeness vs Operational Disruption Trade-Off

Broad scanning improves completeness but may conflict with service stability constraints and limited change windows. Organization A adopted staged scanning aligned to criticality and operational tolerance, prioritizing high-value environments first (e.g., PKI, payment rails, customer identity), then expanding.

## 7 Governance Controls and Evidence Artefacts

Given the observation in Section 5, Organization A was able to devise a challenge resolution analysis to structure the knowledge available within the institution. In this regard, Table 3 summarizes observed challenges, root causes, recommended governance controls, and evidence artefacts that support auditability and repeatability.

**Table 3.** Challenge Resolution Analysis (i.e., Challenges Identification → Root causes listing → Governance control scrutiny → Evidence artefact selection)

| Challenge observed | Root cause | Governance control (recommended) | Evidence artefact (what to keep) |
|---|---|---|---|
| Fragmented crypto ownership ("nobody owns the crypto end-to-end") | Crypto embedded across Dev, Ops, vendors, product lines | Crypto Governance Board + RACI for crypto assets and PQC readiness | Approved RACI matrix; governance minutes; named crypto owners per asset |
| "Hidden crypto" in libraries, middleware, appliances | Non-transparent vendor components and legacy stack complexity | Crypto-SBOM requirement + mandatory crypto disclosure in architecture reviews | Crypto-SBOM templates; architecture sign-offs; vendor crypto declarations |
| Uneven discovery evidence quality | Legacy systems lack telemetry; inconsistent scanning coverage | Evidence confidence scoring + discovery coverage thresholds | Coverage map; confidence score per asset; exception register |
| Third-party PQC uncertainty blocks prioritization | Providers cannot confirm PQC roadmap/dependencies | Supplier PQC readiness clauses + periodic assurance reviews | Contract clauses; supplier questionnaires; readiness attestations |
| Tool scanning disrupts operations or limited by change windows | Production constraints; risk of disruption | Staged scanning aligned to criticality & tolerance | Scanning plan; change tickets; monitoring results |
| Inconsistent certificate lifecycle management | Multiple CAs; decentralized issuance | Centralized certificate governance (PKI policy + inventory ownership) | Certificate inventory export; PKI policy; renewal KPIs; audit logs |
| Prioritization debate ("what goes first?") | Competing priorities; unclear metrics | Exposure register with prioritization logic | Register baseline; prioritization method; executive decision logs |
| Risk model not aligned with retention obligations | Retention horizons not mapped | Longevity classification integrated with crypto inventory | Retention mapping; "long-lived confidentiality" list |
| Over-focus on algorithms instead of systems | PQC treated as "swap RSA" | Crypto-agility architecture standards | Reference architecture; design patterns; engineering standards |
| Lack of accountability for completion | No measurable end-state | PQC readiness metrics integrated into resilience reporting | KPI dashboard; readiness scorecard; milestone evidence pack |

While the case study provides practical and operationally grounded insights into post-quantum cryptographic discovery and governance implementation, several limitations should be acknowledged.
First, the findings are derived from a single large critical service provider operating in a regulated European environment. Although many observed challenges (e.g., cryptographic visibility gaps, ownership fragmentation, and supplier dependency constraints) are likely generalized across large enterprises and financial institutions, organizational structures, regulatory maturity, and technology estate composition may influence the transferability of specific implementation details.
Second, the empirical data presented is anonymized to preserve institutional confidentiality and operational security. While this enables real-world scenario illustration without exposing sensitive infrastructure details, it limits the reproducibility of certain operational metrics and prevents direct external validation of specific asset-level observations.
Third, the discovery methodology is intentionally tooling-vendor neutral. The approach focuses on capability classes (e.g., static cryptographic discovery, runtime telemetry inspection, dependency mapping) rather than specific commercial or open-source tools. While this improves portability across institutions, it may obscure performance or coverage variations that are tool-specific in real deployments.
Fourth, uncertainty remains regarding the precise timeline for cryptographically relevant quantum computing (CRQC). Threat horizon modelling in this work is therefore scenario-based rather than deterministic. As CRQC forecasting remains subject to technological, economic, and geopolitical variability, the exposure modelling logic should be periodically recalibrated as new scientific and intelligence inputs emerge.
Despite these limitations, the case study demonstrates a repeatable governance pattern for converting cryptographic uncertainty into measurable exposure and prioritization artefacts suitable for enterprise risk management and operational resilience planning.

# 8 Quantitative Results and Practical Lessons Learned

The discovery phase, conducted over a 14-month period, provided granular visibility into Organization A's cryptographic ecosystem. The tooling covered 148 heterogeneous environments, scanned 24,350 endpoints, and catalogued 87,200 certificates. The baseline analysis showed continued reliance on classical cryptography, with approximately 68% RSA (primarily legacy) and 32% ECC (primarily newer digital channels).

Four practical lessons emerged that redefined the PQC governance strategy:

**1 Ownership Fragmentation Is the Largest Blocker**

For approximately 40% of cryptographic assets, the operational custodian was unclear (infrastructure teams, application owners, or external vendors). Without a defined owner in a RACI model, migration decisions become ungovernable (downtime approval, testing ownership, rollback responsibility), creating administrative deadlock.

### 2 Evidence Confidence Scoring Is Essential in Legacy Estates

Modern systems provided precise telemetry, while legacy platforms yielded incomplete signals. Introducing a structured evidence confidence score prevented budget misallocation to false positives and reduced the risk of ignoring true exposures in low-visibility systems ("false negatives").

### 3 Third-Party Inheritance Drives Migration Feasibility

Many high-critical services inherited cryptographic primitives from SaaS platforms or vendor payment libraries. In such cases, migration feasibility was effectively constrained by supplier release cycles. The roadmap was therefore restructured to align engineering planning with supplier readiness governance.

### 4 HNDL-Based Prioritisation Identified a Small High-Risk Set

Cross-referencing the inventory with retention and confidentiality obligations prioritized 12 critical services as time-exposed under HNDL risk. While representing under 3% of the overall portfolio, these services contained long-lived data exceeding 10–15 years of confidentiality horizon, making them candidates for accelerated transition.

## 10 Conclusions

PQC readiness is frequently slowed by governance fragmentation, insufficient cryptographic visibility, and supplier dependency complexity rather than algorithm selection alone. The case study demonstrates that theorized temporal exposure, as with Mosca inequalities, modelling can be transformed into actionable governance artefacts supporting PQC transition planning. For organisations operating under "harvest-now, decrypt-later" threat models, discovery is not preparatory overhead but an epistemic precondition for making PQC migration feasible and governable. The case study indicates that structured discovery and evidence confidence scoring are necessary preconditions for prioritising PQC migration in complex environments. Further work is required to validate the approach across additional organisations and sectors.

# 10. References

1. ENISA: Post-quantum cryptography: Current state and quantum mitigation. https://www.enisa.europa.eu/publications/post-quantum-cryptography-current-state-and-quantum-mitigation, last accessed 2026/01/16.
2. Quantum Computing Report: NIST NCCoE Publishes Drafts on Migration to Post-Quantum Cryptography NIST NCCoE Publishes Drafts on Migration to Post-Quantum Cryptography - Quantum Computing Report, last accessed 2026/02/01.
3. ETSI: Quantum-safe cryptography (QSC) programme overview. https://www.etsi.org/technologies/quantum-safe-cryptography, last accessed 2026/01/11.
4. Federal Reserve Board: "Harvest now, decrypt later": Examining post-quantum cryptography and the data privacy risks for distributed ledger networks. FEDS Notes (2025). https://www.federalreserve.gov/econres/notes/feds-notes/harvest-now-decrypt-later-examining-post-quantum-cryptography-and-the-data-privacy-risks-for-distributed-ledger-networks-20250101.html , last accessed 2026/01/16.
5. Cloud Security Alliance: Preparing for the Era of Post-Quantum Cryptography https://cloudsecurityalliance.org/blog/2024/02/14/preparing-for-the-era-of-post-quantum-cryptography , last accessed 2026/01/16.
6. NIST: Post-quantum cryptography. https://csrc.nist.gov/projects/post-quantum-cryptography, last accessed 2026/01/16.
7. Global Risk Institute: Quantum threat timeline report 2024. https://globalriskinstitute.org/publications/quantum-threat-timeline-report-2024/, last accessed 2026/02/10.
8. Mosca, M.: Cybersecurity in an era with quantum computers: Will we be ready? IEEE Security & Privacy 16(5), 38–41 (2018). https://doi.org/10.1109/MSP.2018.3761723.
9. NIST: NIST Releases First 3 Finalized Post-Quantum Encryption Standards https://www.nist.gov/news-events/news/2024/08/nist-releases-first-3-finalized-post-quantum-encryption-standards, last accessed 2026/01/19.
10. IETF: The transport layer security (TLS) protocol version 1.3. RFC 8446 (2018). https://datatracker.ietf.org/doc/rfc8446/, last accessed 2026/01/21.
11. OWASP: Cryptographic storage cheat sheet. https://cheatsheetseries.owasp.org/cheatsheets/Cryptographic_Storage_Cheat_Sheet.html, last accessed 2026/01/21.
12. European Union: Regulation (EU) 2022/2554 of the European Parliament and of the Council on digital operational resilience for the financial sector (DORA) (2022). https://eur-lex.europa.eu/eli/reg/2022/2554/oj, last accessed 2026/01/21.
13. Anonymous: Post-quantum exposure. Unpublished manuscript (2026).
14. NTIA: The minimum elements for a software bill of materials (SBOM) (2021). https://www.ntia.gov/report/2021/minimum-elements-software-bill-materials-sbom, last accessed 2026/02/01.
15. NIST: Secure software development framework (SSDF). https://csrc.nist.gov/projects/ssdf, last accessed 2026/02/03.
16. CISA: SBOM sharing lifecycle report. https://www.cisa.gov/sbom, last accessed 2026/02/03.
17. The Open Group: The Open FAIR™ body of knowledge. https://www.opengroup.org/open-fair, last accessed 2026/02/03.
18. NIST: Guide for conducting risk assessments. NIST SP 800-30 Rev.1 (2012). https://csrc.nist.gov/publications/detail/sp/800-30/rev-1/final , last accessed 2026/02/03.
19. Stebila, D., Fluhrer, S., Gueron, S.: Hybrid key exchange in TLS 1.3. IETF Internet-Draft (2025). https://datatracker.ietf.org/doc/html/draft-ietf-tls-hybrid-design-16, last accessed 2026/02/03.
20. Google: Experimenting with post-quantum cryptography in TLS. Google Security Blog. https://security.googleblog.com/2016/07/experimenting-with-post-quantum.html, last accessed 2026/02/03.